\documentclass[preprint,tightenlines,superscriptaddress,showpacs]{revtex4}

\usepackage{epsfig}
\usepackage{dcolumn}
\usepackage{bm}
\usepackage{amsmath}
\usepackage{color}
\usepackage{subfigure}
\usepackage{graphicx}
\newcommand{\Ra}{R_{z_\alpha}}
\newcommand{\Rb}{R_{z_\beta}}
\newcommand{\Ia}{I_{z_\alpha}}
\newcommand{\Ib}{I_{z_\beta}}
\newcommand{\RF}{R_{F}}
\newcommand{\RH}{R_{H}}
\newcommand{\IF}{I_{F}}
\newcommand{\IH}{I_{H}}%}}}
\newcommand{\BR}{{\cal B}}

\begin{document}
\title{A mathematical solve on the three-interfering-resonances' parameters}

%\title{\boldmath Resonant parameters of the}

\author{X.~Han}
\affiliation{School of Physics and Nuclear Energy Engineering,
Beihang University, Beijing 100191, China}
\author{C.~P.~Shen}
\email{shencp@buaa.edu.cn} \affiliation{School of Physics and
Nuclear Energy Engineering, Beihang University, Beijing 100191,
China}
%\author{C.~Z.~Yuan}
%\affiliation{Institute of High Energy Physics, Chinese Academy of Sciences, Beijing 100049, China}
%\affiliation{University of Chinese Academy of Sciences, Beijing 100049, China}

\begin{abstract}%{{{

The multiple-solution problem in determining the
three-interfering-resonances' parameters from a fit to
an experimentally measured distribution is considered in a
mathematical viewpoint. In this paper it is shown that
there are four numerical solutions for the fit with
three coherent Breit-Wigner functions. Although the explicit analytical
formulae can not be derived in this case, we provide some
constraint equations between the four solutions. For the cases of
nonrelativistic and relativistic Breit-Wigner forms of amplitude
functions, numerical method is provided to derive the other
solutions from the already obtained one based on the obtained constraint equations. In real experimental
measurements with more complicated amplitude forms similar to Breit-Wigner functions, the same method can be
deduced and performed to get numerical solutions. The well agreement between the solved solutions
using this mathematical method and those from the fit directly
verifies the correctness of the supplied constraint equations and mathematical
methodology.
%\noindent \bf{keywords}: multiple solutions; high energy physics experiments.
\end{abstract}%}}}


\maketitle

\section{Introduction}%{{{
\label{sec:i}

One of the main aims during the physics analysis of experimental
data is determination of the parameters of several resonances by
fitting the cross sections or measured mass spectrum with possible
interference between the resonances considered. In some cases,
although the fitted results with interference are not taken as
nominal results, the interference still needs to be considered as
an estimate of the systematic uncertainty.

In particle physics, we usually take Breit-Wigner (BW) function to
represent resonance amplitude. And a typical task
is determination of the BW parameters
from the fit to the measured
distributions in experiment, such as cross sections.
The measured physical quantities are usually in proportion to the modulus of the total
amplitude squared, for examples, $|BW_1+BW_2 e^{i\phi}|^2$ for two
interfering resonances and
$|BW_1+BW_2e^{i\phi_{1}}+BW_3e^{i\phi_{2}|^2}$ for three
interfering resonances, where $\phi$, $\phi_1$, and $\phi_2$ are
the relative phases between resonances. Due to this
square operation in the amplitudes to connect with
the measured physical quantities, we could find
multi-solutions in extracting
amplitudes from the fit to the experimental
measurements. Often it occurs that these
multi-solutions have the same goodness-of-the-fit, and resonance mass
and width, but relative phases are different. This indicates
that different solutions have different coupling strength
to decay channels, which would result in different interpretations
in physics.
Therefore for the fit with interfering resonances,
we need to make sure that all
the solutions have been found.  		
If there are multiple solutions, but only one is reported,
the experimental results may be incomplete or even biased.

Recently, more and more
experimental analyses, especially the studies of the vector
charmonium-like $Y$ states, have indicated this. For example, in
Ref.~\cite{wang} two or three coherent resonances plus
an incoherent background shape are used to fit
the $\pi^+\pi^-\psi(2S)$ invariant mass
distribution. Correspondingly two or four
solutions are found with identical resonance mass and width but
different couplings to electron-positron pairs.  Another example
is presented in Ref.~\cite{yuan2010multiple}, where two solutions
are found in the branching fraction measurement for $\phi \to
\omega \pi^0$ process and the study of $\rho-\omega$ mixing.

In real physics analyses, all the multiple solutions are found via
fitting process. Due to the background statistical fluctuation
or limited statistics, not all the solutions can be found
easily in some cases.  Therefore, from the
mathematical point of view, a nature question is raised: if a
particular solution has been found, then whether other solutions can be
derived from it. For the above question, the authors in
Refs.~\cite{zhu2011mathematical,bukin} proved that if
we use two coherent BW functions to fit the measured
distribution, there should be only two different
solutions, and they can be derived each other by using analytical
formulae and a numerical method. As pointed out in
Ref.~\cite{bukin}, in the case of three resonances with constant
widths there occurred four solutions with the same likelihood
function minimum, but analytical solution of this problem appeared too
hard due to technical difficulties.

In this paper, we discuss the multiple-solution problem in
determining the resonant parameters of three interfering
resonances in a mathematical viewpoint. Although the explicit
analytical formulae can not be derived, we provide some constraint
equations between four solutions. We also provide a mathematical
method to get additional solutions from the obtained one.

This work is organized as follows. After the Introduction, we
present a general mathematic model for the amplitudes of three coherent resonance states
in Sec.~\ref{sec:mm}. If three resonances are described
by the normal BW functions, the analytical expressions for the
relationship between the four solutions are deduced and obtained. An effective
approach is developed to obtain the algebra equations
of the relationship between the four solutions. In Sec.~\ref{sec:fcs}, the relations between
the four solutions are also deduced for relativistic BW forms. In Sec.~\ref{sec:caa},
two numerical examples produced by toy Monte Carlo (MC) are utilized to cross check and confirm our results.
When the form of resonance amplitude is extremely complex,
we can take a similar numerical procedure to obtain other unknown solutions
from the known one. Finally, in Sec.~\ref{sec:d}, a short discussion is
given.
%}}}

\section{Mathematical methodology for three simple-BW-amplitudes case}%{{{
\label{sec:mm}

In the light of two distinct features: (1) all
solutions have the same goodness-of-fit; (2) different solutions have
identical resonance mass and width but different couplings to
electron-positron pairs, we construct a general mathematical model
for multiple solutions based on three interfering amplitude
functions.

A sum of three quantum amplitudes can be described by a complex
function $e(x,z_1,z_2,z_3)$ with form
\begin{equation}
\label{eq:gen}
 e(x,z_1,z_2,z_3) = z_1\ g(x) + z_2\ f(x) + z_3\ h(x)~,
\end{equation}
where $x$ is a measured variable,
$g(x)$,~$h(x)$, and $f(x)$ are complex functions of $x$,
and $z_1$, $z_2$, and $z_3$ are complex numbers. Our
purpose is to find different parameters $z'_1$,
$z'_2$, and $z'_3$ satisfying
\begin{equation}
 \left| e(x,z_1,z_2,z_3) \right|^2 = \left| e(x,z'_1,z'_2,z'_3) \right|^2~.
\label{eq:sol}
\end{equation}
Since the global phase
does not work on amplitude squared operation
we can reduce the dimension of $\{z_1,z_2,z_3\}$
parameter space to a $\{d,z_\alpha,z_\beta\}$ parameter space, where $d$
is a real number. The module of the amplitude
squared of $e(x,z_1,z_2,z_3)$, $\left| e(x,z_1,z_2,z_3) \right|^2$,
can be rewritten in a more convenient form by defining
\begin{eqnarray}
&&\left| e(x,z_1,z_2,z_3) \right|^2 \equiv \frac{1}{d}\left|
g(x) + z_\alpha f(x) + z_\beta h(x) \right|^2 = \frac{\left| g(x) \right|^2}{d} \left| 1+z_\alpha
\frac{f(x)}{g(x)} + z_\beta\frac{h(x)}{g(x)}\right|^2 \cr
&&\equiv \frac{\left| g(x) \right|^2}{d} \left| 1+z_\alpha F(x)+z_\beta H(x) \right|^2
    \equiv  \frac{\left| g(x) \right|^2}{d} E(x,z_\alpha,z_\beta) ~.
\label{eq:eandeprm}
\end{eqnarray}
Here $F(x) \equiv f(x)/g(x)$, $H(x) \equiv h(x)/g(x)$. Considering
$\left| g(x)\right|^2$ is only a product factor and is
independent of $z_{\alpha}$, $z_{\beta}$, and $d$, we
remove it in the following derivation. What we need to
do now is to find different $z_\alpha$, $z_\beta$, and $d$ values which keep
$E(x,z_\alpha,z_\beta)/d$ unchanged.

Taking ($R_F(x)$, $I_F(x)$), ($R_H(x)$, $I_H(x)$), ($\Ra$, $\Ia$), and
($\Rb$, $\Ib$) as real and imaginary parts of $F(x)$, $H(x)$, $z_{\alpha}$,
and $z_{\beta}$, respectively, and using them to represent
$E(x,z_\alpha,z_\beta)$, we get
\begin{align}
  E(x,z_\alpha,z_\beta) = 1
&+ (\RF^2 + \IF^2) (\Ra^2 + \Ia^2) + 2 \Ra \RF - 2 \Ia \IF\nonumber\\
&+ (\RH^2 + \IH^2) (\Rb^2 + \Ib^2) + 2 \Rb \RH - 2 \Ib \IH\nonumber\\
&+ 2 (\RF \RH + \IF \IH) (\Ra \Rb + \Ia \Ib)\nonumber\\
&- 2 (\RF \IH - \IF \RH) (\Ra \Ib - \Ia \Rb).
\end{align}
For the sake of brevity, the specific form of dependence of $R_F(x)$,
$I_F(x)$, $R_H(x)$, and $I_H(x)$ on $x$ is
removed here. Without loss of
generality, we take $d=1$ as an initial solution for convenience.
The next task is to find all the possible $z_\alpha'$, $z_\beta'$, and $d'$ values
to make $E(x,z_\alpha',z_\beta')/d' = E(x,z_\alpha,z_\beta)$.
To be more specific about our work, we consider
that $g(x)$, $h(x)$, and $f(x)$ are widely accepted
nonrelativistic BW functions as an example.
\begin{eqnarray}
\label{eq:nonbw}
    g(x) = \frac{\Gamma_{g}}{(x-M_{g})+i\Gamma_{g}},~~~~f(x) = \frac{\Gamma_{f}}{(x-M_{f})+i\Gamma_{f}},~~~~h(x) = \frac{\Gamma_{h}}{(x-M_{h})+i\Gamma_{h}},
\end{eqnarray}
where $M$ is the mass and $\Gamma$ is the width for a resonance,
respectively.
Using the above forms of $g(x)$, $h(x)$, and $f(x)$,
the real and imaginary parts of $F(x)$ and $H(x)$ become
 $$
 R_F =\frac{\Gamma_f[ \Gamma_g \Gamma_f + (M_g -x)(M_f -x)]}
 {\Gamma_g [\Gamma_f^2 + (M_f-x)^2]},~~~
 I_F =\frac{\Gamma_f[ \Gamma_f (M_g -x) - \Gamma_g (M_f -x)]}
 {\Gamma_g [\Gamma_f^2 + (M_f-x)^2]},
 $$
 $$
 R_H =\frac{\Gamma_h[ \Gamma_g \Gamma_h + (M_g -x)(M_h -x)]}
 {\Gamma_g [\Gamma_h^2 + (M_h-x)^2]},~~~
 I_H =\frac{\Gamma_h[ \Gamma_h (M_g -x) - \Gamma_g (M_h -x)]}
 {\Gamma_g [\Gamma_h^2 + (M_h-x)^2]},
 $$
 respectively.
After some algebra, we obtain the interesting relations below:
\begin{eqnarray}
\label{eq:abcf}
R_F^2 + I_F^2 = a_f R_F + b_f I_F + c_f,~~~
R_H^2 + I_H^2 = a_h R_H + b_h I_H + c_h,
\end{eqnarray}
with
%\textcolor{blue}{
 \begin{eqnarray}
 \label{eq:abc}
a_f = \frac{\Gamma_g + \Gamma_f}{\Gamma_g}\;,\;\;
b_f = \frac{M_g - M_f}{\Gamma_g}\;,\;\;
c_f = -\frac{\Gamma_f}{\Gamma_g}\;,\;\;\\
a_h = \frac{\Gamma_g + \Gamma_h}{\Gamma_g}\;,\;\;
b_h = \frac{M_g - M_h}{\Gamma_g}\;,\;\;
c_h = -\frac{\Gamma_h}{\Gamma_g}\;.\;\;
 \end{eqnarray}
%}
With Eq.~(\ref{eq:abcf}), $E(x,z_\alpha,z_\beta)$ is recast as
\begin{align}
  E(x,z_\alpha,z_\beta) =
~& R_F(a_f\Ra^2 + a_f\Ia^2 + 2\Ra) + I_F(b_f\Ra^2 + b_f\Ia^2 - 2\Ia )\nonumber\\
+& R_H(a_h\Rb^2 + a_h\Ib^2 + 2\Rb) + I_H(b_h\Rb^2 + b_h\Ib^2 - 2\Ib )\nonumber\\
+& 2\ (\RF \RH + \IF \IH) (\Ra \Rb + \Ia \Ib)\\
-& 2\ (\RF \IH - \IF \RH) (\Ra \Ib - \Ia \Rb)\nonumber\\
+& c_f(\Ra^2 + \Ia^2) + c_h(\Rb^2 + \Ib^2) + 1.\nonumber
\end{align}
Similar expression can be obtained for $E(x,z_\alpha',z_\beta')$.
Notice that $R_F$, $I_F$, $R_H$, and $I_H$ are functions in
variable space (namely $x$ space), and $[c_f(\Ra^2 + \Ia^2) +
c_h(\Rb^2 + \Ib^2) + 1]$ is a constant for $x$ space.
We noticed that the term $(\RF \RH + \IF \IH)$
and the linear combination of $R_F$, $I_F$, $R_H$, and $I_H$ have the same number
of $x$ terms with the same power. It is the same for the term $(\RF \IH - \IF \RH)$.
So there are linear correlations for $(\RF
\RH + \IF \IH)$ and $(\RF \IH - \IF \RH)$ by factors
$\{c_1,c_2,c_3,c_4,c_5\}$ and $\{c_6,c_7,c_8,c_9,c_{10}\}$,
respectively. That means $(\RF \RH + \IF \IH)$ and $(\RF \IH - \IF
\RH)$ can be represented by $R_F$, $I_F$, $R_H$, $I_H$, and a
constant term.
\begin{eqnarray}
\label{eq:lirel}
\RF \RH + \IF \IH = c_1 R_F + c_2 I_F + c_3 R_H + c_4 I_H +    c_5[c_f(\Ra^2 + \Ia^2) + c_h(\Rb^2 + \Ib^2) + 1], \\
\RF \IH - \IF \RH = c_6 R_F + c_7 I_F + c_8 R_H + c_9 I_H + c_{10}[c_f(\Ra^2 + \Ia^2) + c_h(\Rb^2 + \Ib^2) + 1]. \nonumber
\end{eqnarray}
The factors $\{c_1,c_2,c_3,c_4,c_5\}$ and $\{c_6,c_7,c_8,c_9,c_{10}\}$ follow Eq.~(\ref{eq:factors}):
\begin{align}%{{{
\label{eq:factors}
c_1 =& \frac{\Gamma_h (M_f^2 + M_g M_h - M_f (M_g + M_h) + (\Gamma_f + \Gamma_g) (\Gamma_f + \Gamma_h))}{\Gamma_g (M_f^2 - 2 M_f M_h + M_h^2 + (\Gamma_f + \Gamma_h)^2)},\nonumber\\
c_2 =& \frac{\Gamma_h (-M_h (\Gamma_f + \Gamma_g) + M_f (\Gamma_g - \Gamma_h) + M_g (\Gamma_f + \Gamma_h))}{\Gamma_g (M_f^2 - 2 M_f M_h + M_h^2 + (\Gamma_f + \Gamma_h)^2)},\nonumber\\
c_3 =& \frac{\Gamma_f (M_f (M_g - M_h) - M_g M_h + M_h^2 + \Gamma_f \Gamma_g + \Gamma_f \Gamma_h + \Gamma_g \Gamma_h + \Gamma_h^2)}{\Gamma_g (M_f^2 - 2 M_f M_h + M_h^2 + (\Gamma_f + \Gamma_h)^2)},\nonumber\\
c_4 =& \frac{\Gamma_f (M_h (-\Gamma_f + \Gamma_g) + M_g (\Gamma_f + \Gamma_h) - M_f (\Gamma_g + \Gamma_h))}{\Gamma_g (M_f^2 - 2 M_f M_h + M_h^2 + (\Gamma_f + \Gamma_h)^2)},\nonumber\\
c_5 =& -\frac{2 \Gamma_f \Gamma_h (\Gamma_f + \Gamma_h)}{\Gamma_g (M_f^2 - 2 M_f M_h + M_h^2 + (\Gamma_f + \Gamma_h)^2)},\\
c_6 =&\frac{\Gamma_h (-M_h (\Gamma_f + \Gamma_g) + M_f (\Gamma_g - \Gamma_h) + M_g (\Gamma_f + \Gamma_h))}{\Gamma_g (M_f^2 - 2 M_f M_h + M_h^2 + (\Gamma_f + \Gamma_h)^2)},\nonumber\\
c_7 =&-\frac{\Gamma_h (M_f^2 + M_g M_h - M_f (M_g + M_h) + (\Gamma_f + \Gamma_g) (\Gamma_f + \Gamma_h))}{\Gamma_g (M_f^2 - 2 M_f M_h + M_h^2 + (\Gamma_f + \Gamma_h)^2)},\nonumber\\
c_8 =& \frac{\Gamma_f (M_h (\Gamma_f - \Gamma_g) - M_g (\Gamma_f + \Gamma_h) + M_f (\Gamma_g + \Gamma_h))}{\Gamma_g (M_f^2 - 2 M_f M_h + M_h^2 + (\Gamma_f + \Gamma_h)^2)},\nonumber\\
c_9 =& \frac{\Gamma_f (M_f (M_g - M_h) - M_g M_h + M_h^2 + \Gamma_f \Gamma_g + \Gamma_f \Gamma_h + \Gamma_g \Gamma_h + \Gamma_h^2)}{\Gamma_g (M_f^2 - 2 M_f M_h + M_h^2 + (\Gamma_f + \Gamma_h)^2)},\nonumber\\
c_{10}=& \frac{2 (-M_f + M_h) \Gamma_f \Gamma_h}{\Gamma_g (M_f^2 - 2 M_f M_h + M_h^2 + (\Gamma_f + \Gamma_h)^2)}.\nonumber
\end{align}%}}}
Then we can get
\begin{align}
\label{eq:dbw}
  E(x,z_\alpha,z_\beta) =
~& R_F(a_f\Ra^2 + a_f\Ia^2 + 2\Ra + c_1 A + c_6 B) + I_F(b_f\Ra^2 + b_f\Ia^2 - 2\Ia + c_2 A + c_7 B )\nonumber\\
+& R_H(a_h\Rb^2 + a_h\Ib^2 + 2\Rb + c_3 A + c_8 B) + I_H(b_h\Rb^2 + b_h\Ib^2 - 2\Ib + c_4 A + c_9 B )\nonumber\\
+& c_f\Ra^2 + c_f\Ia^2+ c_h\Rb^2 + c_h\Ib^2 + 1 + c_5 A + c_{10} B,
\end{align}
with
$ A =2(\Ra \Rb + \Ia \Ib) $ and
$ B =-2(\Ra \Ib - \Ia \Rb) $.

We know that $\Ra$, $\Ia$, $\Rb$, and $\Ib$ are functions in
parameter space $\{d,z_\alpha,z_\beta\}$. If we want to make
$E(x,z_\alpha',z_\beta')/d' = E(x,z_\alpha,z_\beta)$ hold for any $x$,
then the corresponding coefficients of the
functions in parameter space should be equal,
which immediately leads to the following equations:
\begin{align}
\label{eq:bwary}
\frac{1}{d'}(a_fR_{z_\alpha}'^2 + a_fI_{z_\alpha}'^2 + 2R_{z_\alpha}' + c_1 A' + c_6 B')&= a_f\Ra^2 + a_f\Ia^2 + 2\Ra + c_1 A + c_6 B,\nonumber \\
\frac{1}{d'}(b_fR_{z_\alpha}'^2 + b_fI_{z_\alpha}'^2 - 2I_{z_\alpha}' + c_2 A' + c_7 B')&= b_f\Ra^2 + b_f\Ia^2 - 2\Ia + c_2 A + c_7 B,\nonumber\\
\frac{1}{d'}(a_hR_{z_\beta }'^2 + a_hI_{z_\beta }'^2 + 2R_{z_\beta }' + c_3 A' + c_8 B')&= a_h\Rb^2 + a_h\Ib^2 + 2\Rb + c_3 A + c_8 B,\\
\frac{1}{d'}(b_hR_{z_\beta }'^2 + b_hI_{z_\beta }'^2 - 2I_{z_\beta }' + c_4 A' + c_9 B')&= b_h\Rb^2 + b_h\Ib^2 - 2\Ib + c_4 A + c_9 B,\nonumber\\
\hspace{-15mm}
\frac{1}{d'}(c_fR_{z_\alpha}'^2 + c_fI_{z_\alpha}'^2 + c_hR_{z_\beta}'^2 + c_hI_{z_\beta}'^2 + 1 + c_5 A' + c_{10} B')&= c_f\Ra^2 + c_f\Ia^2+ c_h\Rb^2 + c_h\Ib^2 + 1 + c_5 A + c_{10} B, \nonumber
\end{align}
with
\begin{eqnarray}
A'=2(\Ra' \Rb' + \Ia' \Ib'),& &
A =2(\Ra \Rb + \Ia \Ib),\nonumber\\
B'=-2(\Ra' \Ib' - \Ia' \Rb'),& &
B =-2(\Ra \Ib - \Ia \Rb).\nonumber
\end{eqnarray}
All what we need is to solve the Eq.~(\ref{eq:bwary}) to obtain
the values of $\Ra'$, $\Ia'$, $\Rb'$, $\Ib'$, and $d'$.
Unfortunately, there are no explicit analytical expressions for
them.
So we can not prove there must be four solutions.
Such conclusion agrees with that in Ref.~\cite{bukin}.
However, by using mathematica tool~\cite{mathe} to input Eq.~(\ref{eq:bwary})
and initial solution, we exactly get four numerical solutions quickly.
The numerical solutions can be taken as cross checks and references compared with those from the fits.
This definitely saves a lot of time and energy.

%In particle physical analyses, usually we use $r\cdot e^{i \phi}$
%instead of $z$, which can be easily transformed by
%\begin{eqnarray}
%\label{ztophi}
%\phi=arg(z)\;,\;\;
%r=|z|.
%\end{eqnarray}
We need to point out that the Eqs.~(\ref{eq:abcf}),
(\ref{eq:lirel}), (\ref{eq:dbw}), and (\ref{eq:bwary}) are
independent on the explicit expressions of BW functions, while the
factors such as $a_f$, $b_f$, $c_f$, $a_h$, $b_h$, $c_h$,
$\{c_1,c_2,c_3,c_4,c_5\}$, and $\{c_6,c_7,c_8,c_9,c_{10}\}$ are
dependent.
%We still can solve  $a_f$, $b_f$, $c_f$, $a_h$, $b_h$ and $c_h$
%by Eq.\ref{eq:abcf} and
%$\{c_1,c_2,c_3,c_4,c_5\}$, $\{c_6,c_7,c_8,c_9,c_{10}\}$
%by Eq.\ref{eq:lirel}.

%}}}

\section{Mathematical Methodology for three relativistic-BW-amplitudes case}%{{{
\label{sec:fcs}
Here we take another form for
$f(x)$ , $g(x)$, and $h(x)$, \textit{i.e.}, relativistic BW
amplitudes that are usually used in $e^+e^-$ reactions to extract
the parameters of $Y$ resonance:
 \begin{equation}
 \label{eq:rbw}
 BW(s)=
 \frac{\sqrt{12\pi \Gamma_{e^+e^-}^R
 \BR_R\Gamma_{R}} }
 {s-M^2_R+i M_R\Gamma_{R}} \sqrt{\frac{PS(\sqrt{s})}{PS(M_R)}},
 \end{equation}
where $s$ is the $e^+e^-$ center-of-mass square; $M_R$ is the mass
of the resonance $R$; $\Gamma_R$ and $\Gamma_{e^+e^-}^R$ are the
total width and partial width to $e^+e^-$, respectively; $\BR_R$
is the branching fraction of the resonance $R$ decays into a final
state; and $PS$ is the $n-$body decay phase space factor which
increases smoothly from the mass threshold with the
$\sqrt{s}$~\cite{pdg}. Notice that the Eq.~(\ref{eq:bwary}) is
independent on the forms of amplitudes, while its coefficients
will change. With some algebra, we can obtain the coefficients for
other forms of amplitudes.

With Eq.~(\ref{eq:rbw}), the $F(x)$ and $H(x) $ are changed to
\begin{eqnarray}
F(x)=
 \frac{x^2-M_g^2+i M_g\Gamma_{g}}
 {x^2-M_f^2+i M_f\Gamma_{f}}
 \sqrt{\frac{\Gamma_{f}PS(M_g)}
 { \Gamma_{g}PS(M_f)}},\nonumber \\
H(x)=
 \frac{x^2-M_g^2+i M_g\Gamma_{g}}
 {x^2-M_h^2+i M_h\Gamma_{h}}
 \sqrt{\frac{ \Gamma_{h}PS(M_g)}
 { \Gamma_{g}PS(M_h)}}.\nonumber
\end{eqnarray}
In this situation, $ \RF $, $ \IF $, $ \RH $, and $ \IH $ are
changed. So we need resolve the parameters $a_f$, $b_f$, $c_f$,
$a_h$, $b_h$, $c_h$, $\{c_1,c_2,c_3,c_4,c_5\}$, and
$\{c_6,c_7,c_8,c_9,c_{10}\}$ using Eqs.~(\ref{eq:abcf}) and
(\ref{eq:lirel}), respectively.
 And we obtain
\begin{eqnarray}%{{{
\label{eq:rwabc}
a_{f}= \frac{(\Gamma_f M_f+\Gamma_g M_g)}{M_f\sqrt{\Gamma_f \Gamma_g}}\sqrt{\frac{PS(M_g)}{PS(M_f)}},\;\;
b_{f}= -\frac{ (M_f^2-M_g^2)}{M_f\sqrt{\Gamma_f \Gamma_g }}\sqrt{\frac{PS(M_g)}{PS(M_f)}}\;,\;\;
c_{f}= -\frac{M_g PS(M_g)}{M_f PS(M_f)}\;,\;\\
a_{h}= \frac{(\Gamma_h M_h+\Gamma_g M_g)}{M_h\sqrt{\Gamma_h \Gamma_g}}\sqrt{\frac{PS(M_g)}{PS(M_h)}}\;,\;\;
b_{h}= -\frac{ (M_h^2-M_g^2)}{M_h\sqrt{\Gamma_h \Gamma_g }}\sqrt{\frac{PS(M_g)}{PS(M_h)}}\;,\;\;
c_{h}= -\frac{M_g PS(M_g)}{M_h PS(M_h)}\;,\;\;
\end{eqnarray}%}}}
and
\begin{align}%{{{
\label{eq:rw-factors}
    \nonumber c_1=& \frac{\sqrt{\Gamma_h  PS(g)} \left[M_f^4-M_f^2 \left(-\Gamma_f^2+M_g^2+M_h^2\right)+\Gamma_f M_f (\Gamma_g M_g+\Gamma_h M_h)+M_g M_h (\Gamma_g \Gamma_h+M_g M_h)\right]}{\sqrt{\Gamma_g  PS(h)} \left[M_f^4+M_f^2 \left(\Gamma_f^2-2 M_h^2\right)+2 \Gamma_f \Gamma_h M_f M_h+M_h^2 \left(\Gamma_h^2+M_h^2\right)\right]},\\
    \nonumber c_2=& -\frac{\sqrt{\Gamma_h  PS(g)} \left[M_f^2 (\Gamma_h M_h-\Gamma_g M_g)+\Gamma_f M_f \left(M_h^2-M_g^2\right)+M_g M_h (\Gamma_g M_h-\Gamma_h M_g)\right]}{\sqrt{\Gamma_g  PS(h)} \left[M_f^4+M_f^2 \left(\Gamma_f^2-2 M_h^2\right)+2 \Gamma_f \Gamma_h M_f M_h+M_h^2 \left(\Gamma_h^2+M_h^2\right)\right]},\\
    \nonumber c_3=&\frac{\sqrt{\Gamma_f  PS(g)} \left[M_f^2 \left(M_g^2-M_h^2\right)+\Gamma_f M_f (\Gamma_g M_g+\Gamma_h M_h)+M_h \left(-M_g^2 M_h+\Gamma_g \Gamma_h M_g+M_h^3+\Gamma_h^2 M_h\right)\right]}{\sqrt{\Gamma_g  PS(f)} \left[M_f^4+M_f^2 \left(\Gamma_f^2-2 M_h^2\right)+2 \Gamma_f \Gamma_h M_f M_h+M_h^2 \left(\Gamma_h^2+M_h^2\right)\right]},\\
    \nonumber c_4=&\frac{\sqrt{\Gamma_f  PS(g)} \left[M_f^2 (-(\Gamma_g M_g+\Gamma_h M_h))+\Gamma_f M_f \left(M_g^2-M_h^2\right)+M_g M_h (\Gamma_h M_g+\Gamma_g M_h)\right]}{\sqrt{\Gamma_g  PS(f)} \left[M_f^4+M_f^2 \left(\Gamma_f^2-2 M_h^2\right)+2 \Gamma_f \Gamma_h M_f M_h+M_h^2 \left(\Gamma_h^2+M_h^2\right)\right]},\\
    c_5=&-\frac{2 M_g PS(g) \sqrt{\Gamma_f \Gamma_h} (\Gamma_f M_f+\Gamma_h M_h)}{\sqrt{PS(f)PS(h)} \left[M_f^4+M_f^2 \left(\Gamma_f^2-2 M_h^2\right)+2 \Gamma_f \Gamma_h M_f M_h+M_h^2 \left(\Gamma_h^2+M_h^2\right)\right]},\\
    \nonumber c_6=&-\frac{\sqrt{\Gamma_h  PS(g)} \left[M_f^2 (\Gamma_h M_h-\Gamma_g M_g)+\Gamma_f M_f \left(M_h^2-M_g^2\right)+M_g M_h (\Gamma_g M_h-\Gamma_h M_g)\right]}{\sqrt{\Gamma_g  PS(h)} \left[M_f^4+M_f^2 \left(\Gamma_f^2-2 M_h^2\right)+2 \Gamma_f \Gamma_h M_f M_h+M_h^2 \left(\Gamma_h^2+M_h^2\right)\right]},\\
    \nonumber c_7=&-\frac{\sqrt{\Gamma_h  PS(g)} \left[M_f^4-M_f^2 \left(-\Gamma_f^2+M_g^2+M_h^2\right)+\Gamma_f M_f (\Gamma_g M_g+\Gamma_h M_h)+M_g M_h (\Gamma_g \Gamma_h+M_g M_h)\right]}{\sqrt{\Gamma_g  PS(h)} \left[M_f^4+M_f^2 \left(\Gamma_f^2-2 M_h^2\right)+2 \Gamma_f \Gamma_h M_f M_h+M_h^2 \left(\Gamma_h^2+M_h^2\right)\right]},\\
    \nonumber c_8=&-\frac{\sqrt{\Gamma_f  PS(g)} \left[M_f^2 (-(\Gamma_g M_g+\Gamma_h M_h))+\Gamma_f M_f \left(M_g^2-M_h^2\right)+M_g M_h (\Gamma_h M_g+\Gamma_g M_h)\right]}{\sqrt{\Gamma_g  PS(f)} \left[M_f^4+M_f^2 \left(\Gamma_f^2-2 M_h^2\right)+2 \Gamma_f \Gamma_h M_f M_h+M_h^2 \left(\Gamma_h^2+M_h^2\right)\right]},\\
    \nonumber c_9=&\frac{\sqrt{\Gamma_f  PS(g)} \left[M_f^2 \left(M_g^2-M_h^2\right)+\Gamma_f M_f (\Gamma_g M_g+\Gamma_h M_h)+M_h \left(-M_g^2 M_h+\Gamma_g \Gamma_h M_g+M_h^3+\Gamma_h^2 M_h\right)\right]}{\sqrt{\Gamma_g  PS(f)} \left[M_f^4+M_f^2 \left(\Gamma_f^2-2 M_h^2\right)+2 \Gamma_f \Gamma_h M_f M_h+M_h^2 \left(\Gamma_h^2+M_h^2\right)\right]},\\
    \nonumber c_{10}=&-\frac{2 M_g PS(g) \sqrt{\Gamma_f \Gamma_h} \left(M_f^2-M_h^2\right)}{\sqrt{PS(f) PS(h)} \left[M_f^4+M_f^2 \left(\Gamma_f^2-2 M_h^2\right)+2 \Gamma_f \Gamma_h M_f M_h+M_h^2 \left(\Gamma_h^2+M_h^2\right)\right]}.
\end{align}%}}}
Substitute the above factors  into Eq.~(\ref{eq:bwary}), the
relationship between multi-solutions can be obtained, therefore,
one can derive the other three solutions from the already obtained
one~\cite{mathe}.
%}}}

\section{Check and Application}%{{{
\label{sec:caa}

\subsection{Simple BW amplitudes}%{{{
\label{secIII}

\begin{figure}[ht]
\includegraphics[scale=.35]{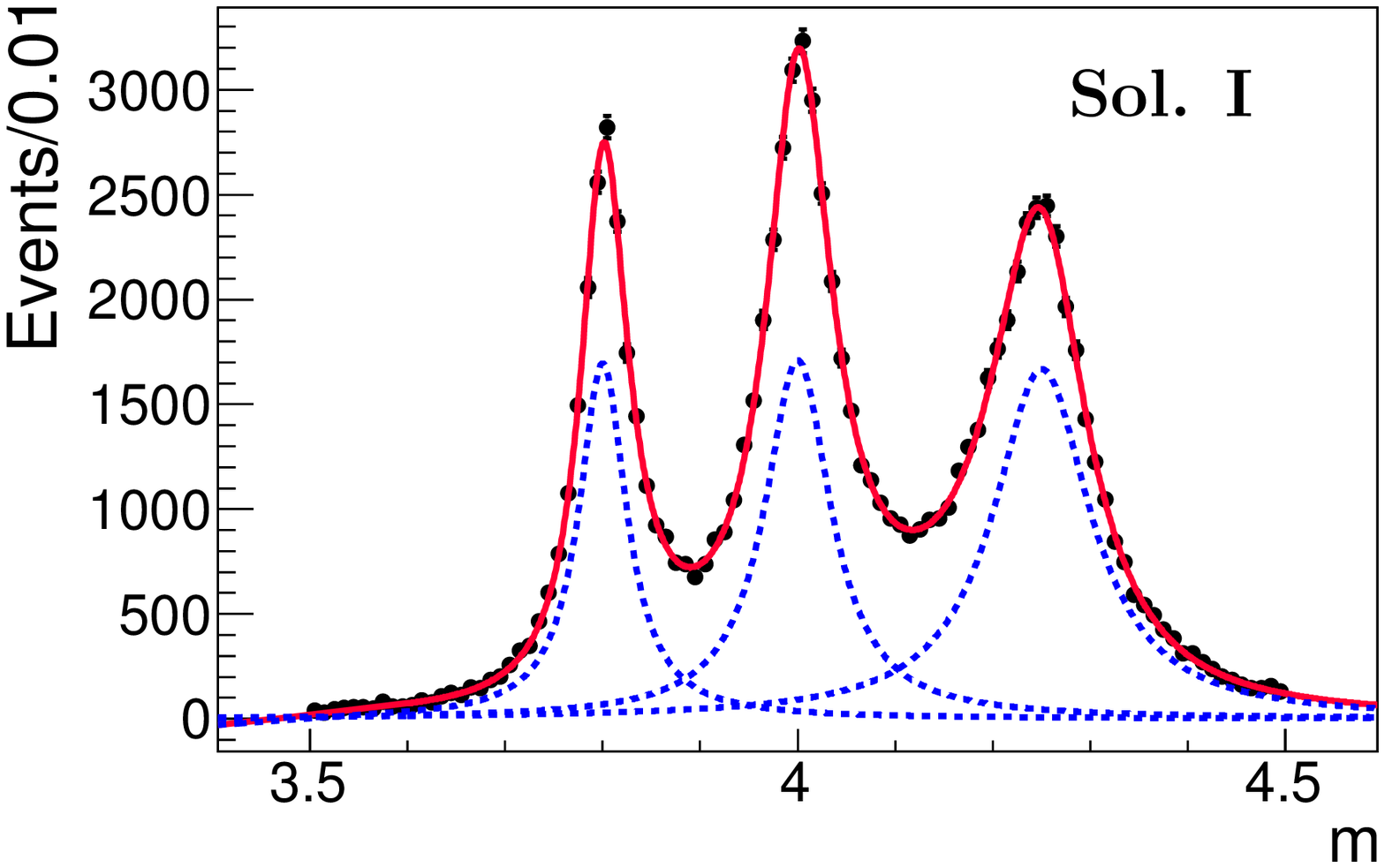}
\includegraphics[scale=.35]{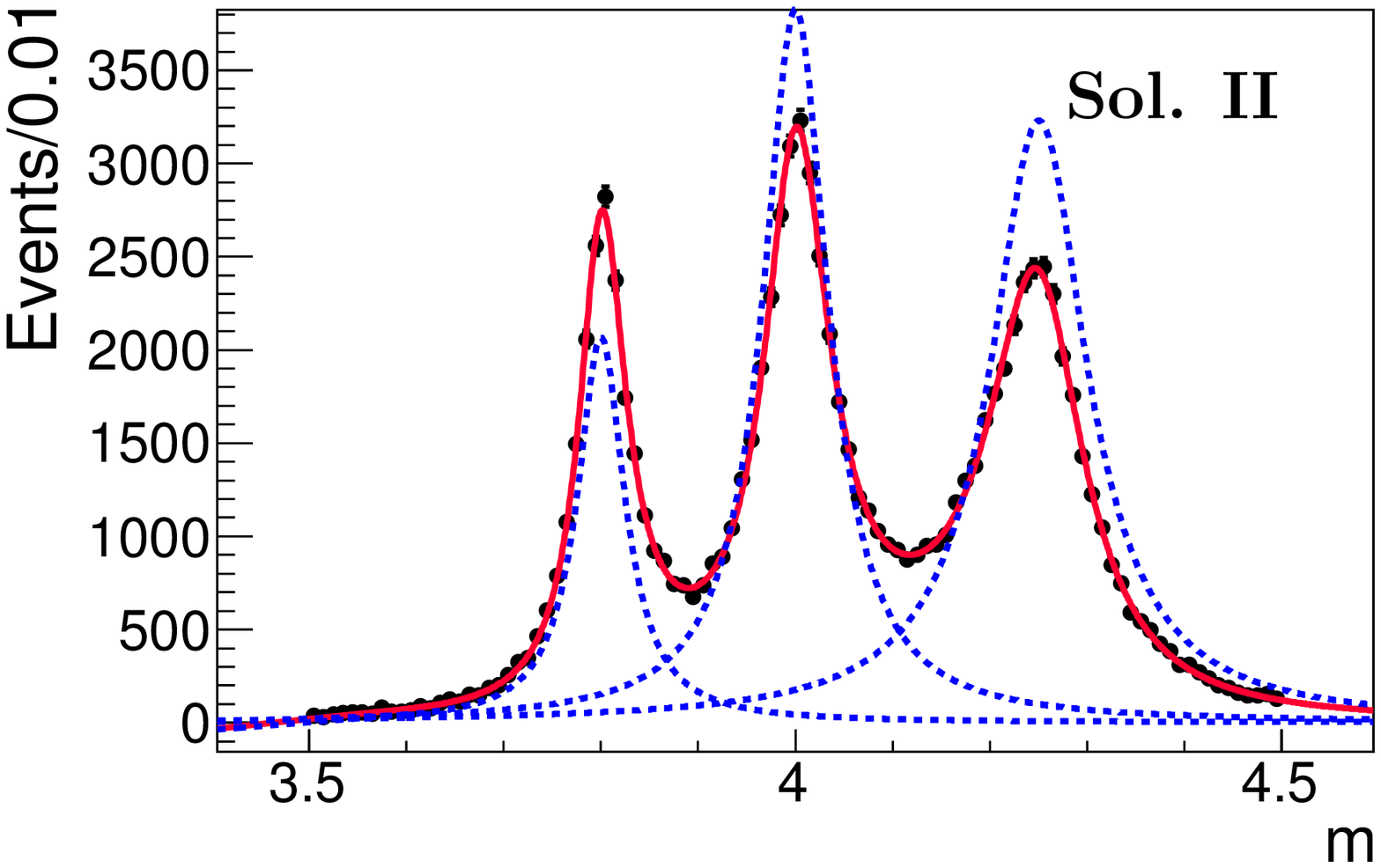}\\
\includegraphics[scale=.35]{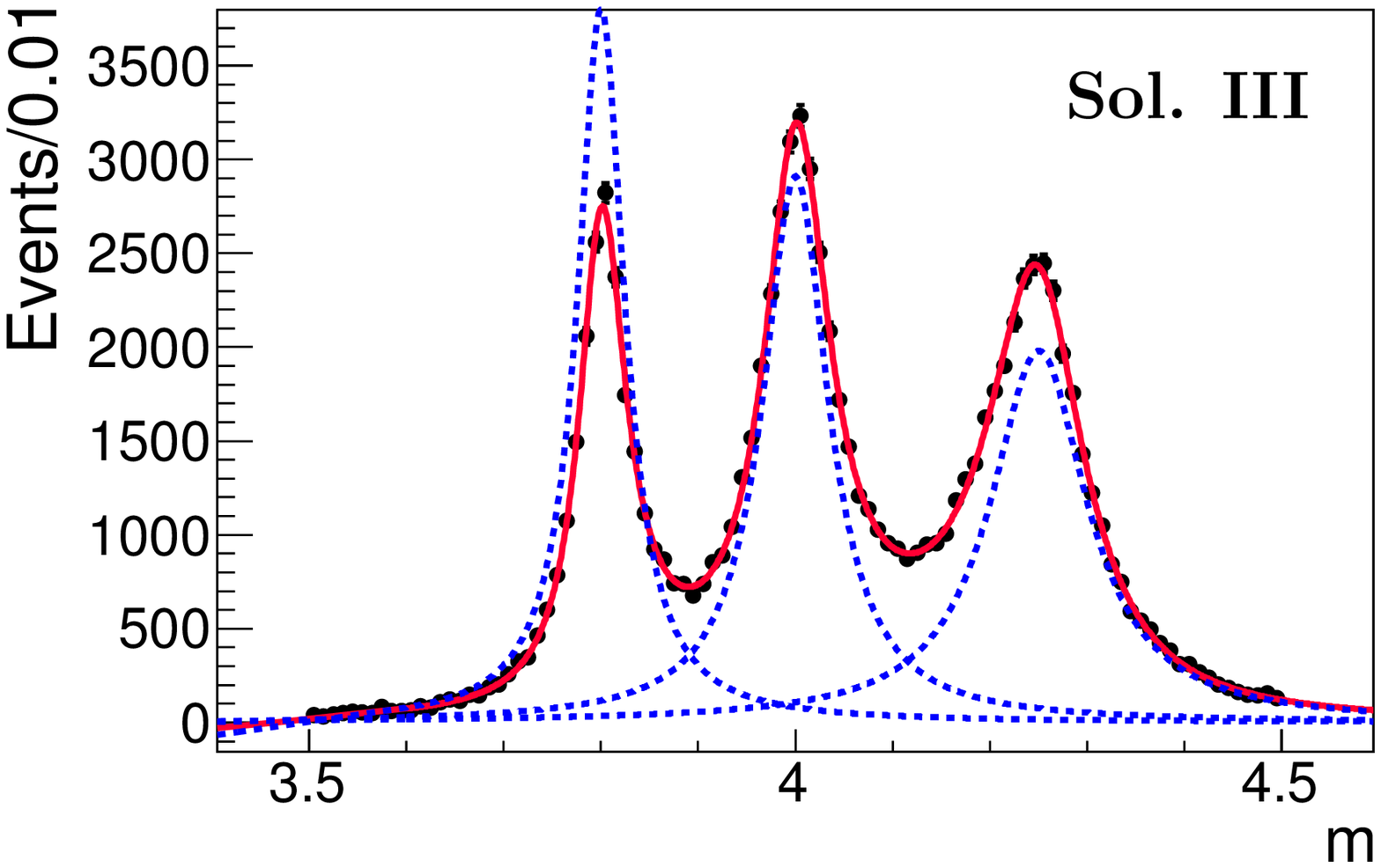}
\includegraphics[scale=.35]{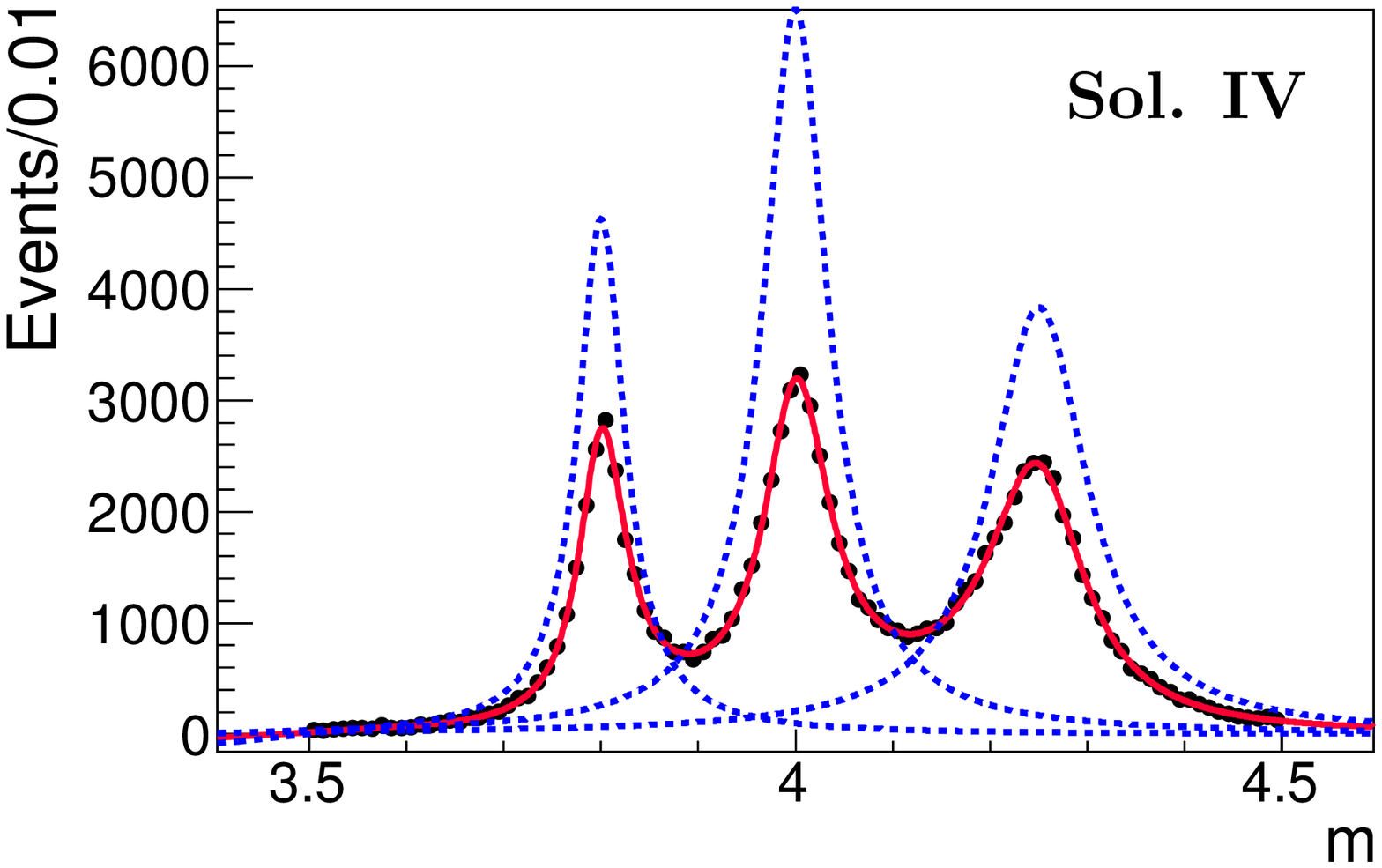}
\caption{The four solutions from the fit to the toy MC produced
mass spectra with the three interfering resonances included. The
solid curves show the best fit and the dashed curves show the
contributions from the three nonrelativistic BW components.}
\label{fig1}
\end{figure}

In order to verify our deduction on constraint equations and
mathematical program in obtaining numerical solutions,
let us take a random example for the case of three
simple BW amplitudes with interference.
The parameter values of the three BW functions as one solution are set as
\begin{eqnarray}
\nonumber & &M_g = 3.80,\;\; \Gamma_g = 0.03,\\
\nonumber & &M_f=4.00,\;\; \Gamma_f=0.04, \;\; \phi_f=\pi/3,\\
\nonumber & &M_h=4.25,\;\;\Gamma_h=0.06,\;\; \phi_h=3\pi/4.
  \end{eqnarray}
%In particle physical analyses, usually we use $r\cdot e^{i \phi}$ instead of $z$,
%which can be easily transformed by
%$ \phi=arg(z) $ and
%$ r=|z| $.
The module of the amplitude
squared of three interfering resonances is $ \left|
BW_{g}(m)+BW_{f}(m)e^{i\phi_{f}}  +
BW_{h}(m)e^{i\phi_{h}}\right|^{2} $ and the BW amplitudes use the
formats shown in Eq.~(\ref{eq:nonbw}).
That is to say $z_{\alpha}=e^{i\phi_f}=1/2+\sqrt{3}/2i$ and
$z_{\beta}=e^{i\phi_h}=-1/\sqrt{2}+1/\sqrt{2}i$ for the above solution.
According to the above
probability density function and the first set of input solution,
toy MC is used to generate a data sample of 100,000
events. The generated distributions with dots with error bars are shown in
Fig.~\ref{fig1}.
An binned extended maximum likelihood fit is applied to
such distribution with three interfering resonances
to extract the parameters of resonances. Four sets of solutions
are found. The fitted results are summarized in Table~\ref{tab:example1}
and the corresponding fitted plots are shown in
Fig.~\ref{fig1} in solid lines.
Using the aforementioned method, we can also obtain another three
sets of solutions numerically. We found the numerical solutions are
exactly repeated by fitting.
For those with little difference, they are consistent within $0.5\sigma$,
where $\sigma$ is the error from the fit.
The comparison of the results is shown
in Table~\ref{tab:example1}.

\begin{table}[htbp]%{{{
\begin{center}
\hspace{-15mm} \caption[Example1]{
Comparison between the extracted
solution using mathematical method and that obtained from the fit
with three interfering simple BW functions.
A data sample of 100,000 events generated by toy MC
is used in the fit.}
\vspace{0.2 cm} \label{tab:example1}
\begin{tabular}{c|cc|cc|cc|cc}
\hline
 Item       &  Sol. I (Input) & Fit I   & Sol. II & Fit II  & Sol. III & Fit III & Sol. IV  & Fit IV  \\
\hline
$\phi_f$   &  $\pi/3$ & 1.06 & 2.29 & 2.29 & 3.56 & 3.55 & 4.79 & 4.80 \\
$\phi_h$   & $3\pi/4$ & 2.37 & 6.02 & 6.02 & 5.66 & 5.67 & 3.05 & 3.05 \\
\hline
$d$        &1         & ---  & 0.81 & ---  & 0.46 & ---  & 0.37 & ---  \\
$\Ra$      &1/2       & 0.50 &-0.89 &-0.89 &-0.81 &-0.81 & 0.10 & 0.10 \\
$\Ia$   &$\sqrt{3}/2$ & 0.87 & 1.02 & 1.02 &-0.36 &-0.35 &-1.19 &-1.17 \\
$\Rb$  &-$\sqrt{2}/2$ &-0.72 & 1.20 & 1.19 & 0.60 & 0.60 &-0.91 &-0.91 \\
$\Ib$   &$\sqrt{2}/2$ & 0.69 &-0.32 &-0.32 &-0.43 &-0.42 & 0.09 & 0.09 \\
\hline
$M_g$      &    3.80  & 3.80 & 3.80 & 3.80 & 3.80 & 3.80 & 3.80 & 3.80 \\
$\Gamma_g$ &    0.03  & 0.03 & 0.03 & 0.03 & 0.03 & 0.03 & 0.03 & 0.03 \\
$M_f$      &    4.00  & 4.00 & 4.00 & 4.00 & 4.00 & 4.00 & 4.00 & 4.00 \\
$\Gamma_f$ &    0.04  & 0.04 & 0.04 & 0.04 & 0.04 & 0.04 & 0.04 & 0.04 \\
$M_h$      &    4.25  & 4.25 & 4.25 & 4.25 & 4.25 & 4.25 & 4.25 & 4.25 \\
$\Gamma_h$ &    0.06  & 0.06 & 0.06 & 0.06 & 0.06 & 0.06 & 0.06 & 0.06 \\
\hline
\end{tabular}
\end{center}
\end{table}%}}}

It is obvious that, for the case of three nonrelativistic BW
amplitudes with interference, if one solution is known from the fit,
the other three can be derived readily and numerically by solving
Eq.~(\ref{eq:bwary}). %}}}

\subsection{Relativistic BW amplitudes}%{{{

For the case of relativistic BW amplitudes with interference,
the values of the parameters as one solution are set as
\begin{eqnarray}
\nonumber & &M_g = 4.20,\;\; \Gamma_g = 0.09 ,\\
\nonumber & &M_f = 4.40,\;\; \Gamma_f=0.12, \;\; \phi_f=\pi/2,\\
\nonumber & &M_h = 4.60,\;\;\Gamma_h=0.18,\;\; \phi_h=3\pi/4.
\end{eqnarray}

The module of the amplitude
squared of three interfering resonances is $ \left|
BW_{g}(m)+BW_{f}(m)e^{i\phi_{f}}  +
BW_{h}(m)e^{i\phi_{h}}\right|^{2} $ and the BW amplitudes use the
formats shown in Eq.~(\ref{eq:rbw}), where for the phase space
factor we assume the reaction process is
$e^{+}e^{-}\to\pi^{+}\pi^{-}J/\psi$.
That is to say $z_{\alpha}=
\sqrt{B_f\Gamma_{e^+e^-}^f/
	       B_g\Gamma_{e^+e^-}^g}e^{i\phi_f}=i$ and
$z_{\beta}=
\sqrt{B_h\Gamma_{e^+e^-}^h/
	       B_g\Gamma_{e^+e^-}^g}e^{i\phi_h}=-1/\sqrt{2}+1/\sqrt{2}i$
for the above solution, where the values of $B_{R}\Gamma^{R}_{e^{+}e^{-}}$
are set as 1 for $R=g,~f,$ and $h$.

\begin{figure}[ht]
\includegraphics[scale=.35]{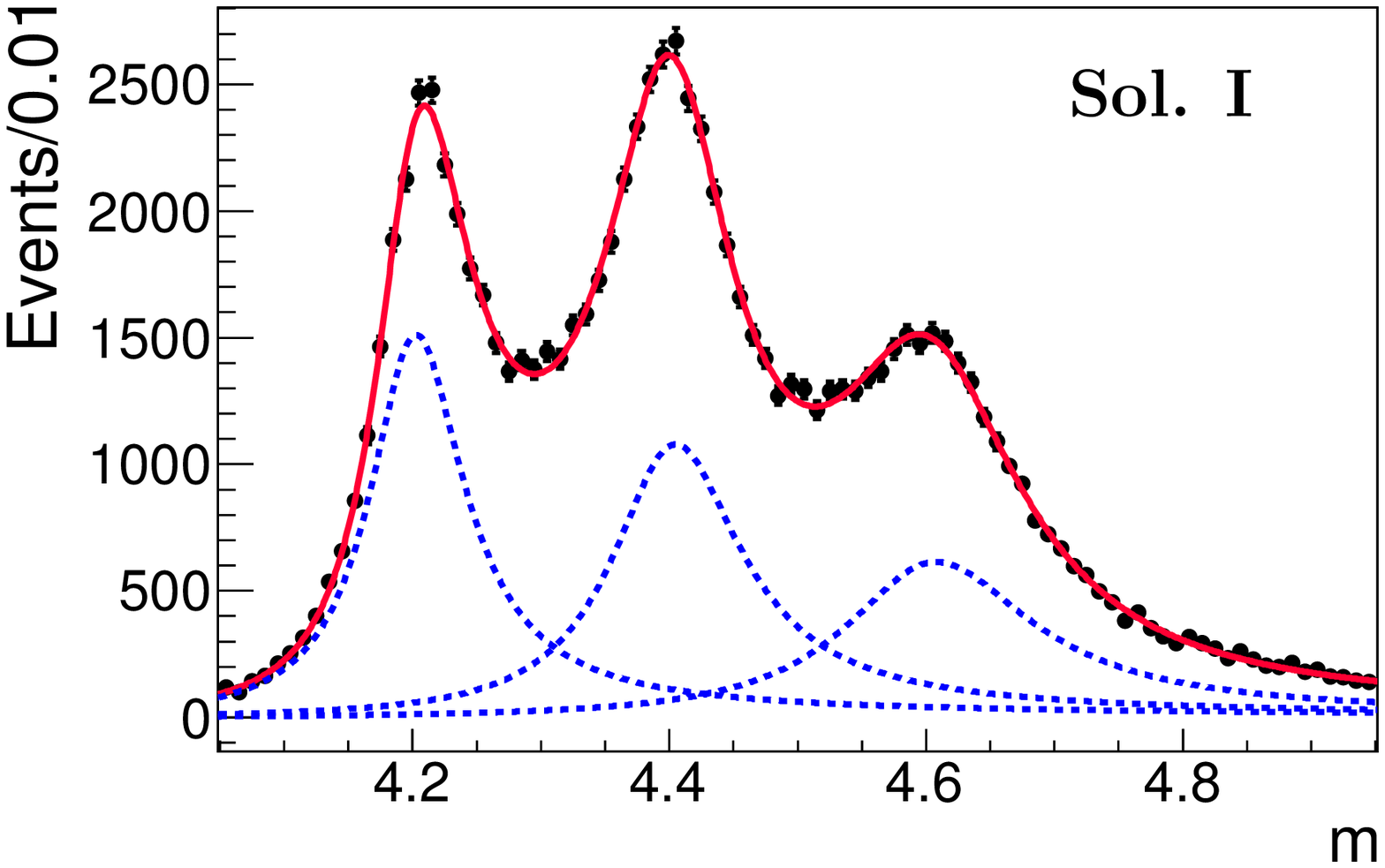}
\includegraphics[scale=.35]{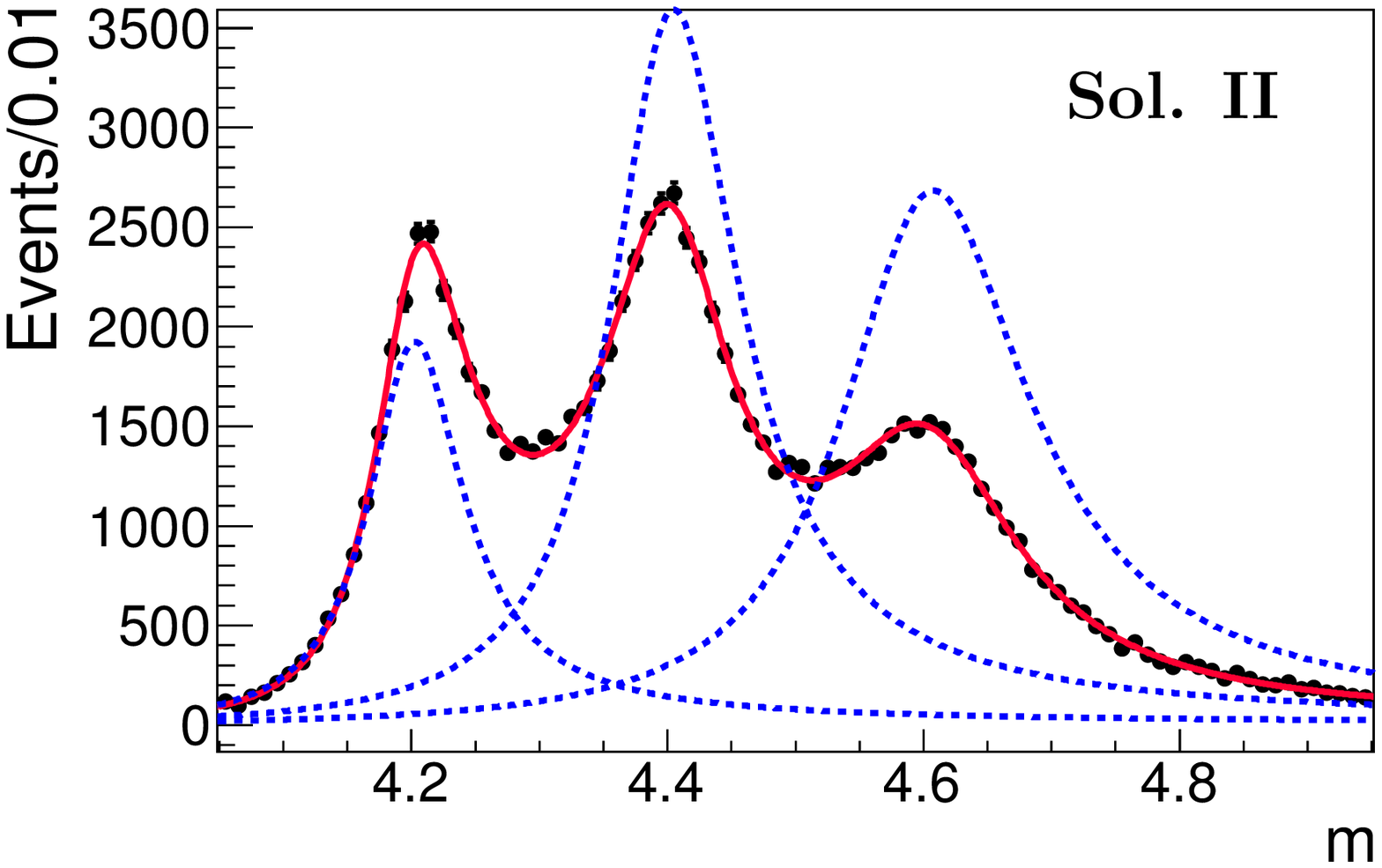}\\
\includegraphics[scale=.35]{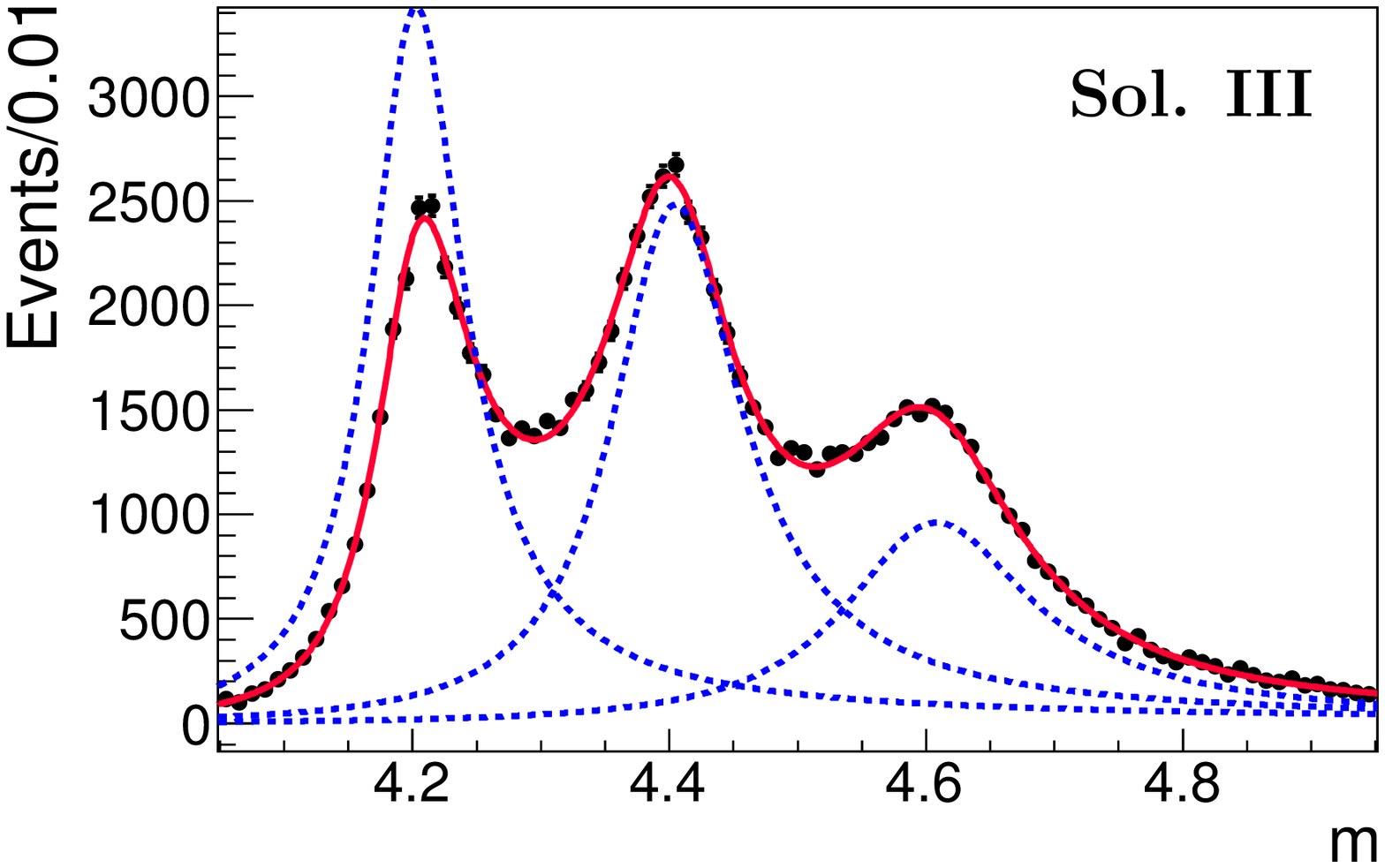}
\includegraphics[scale=.35]{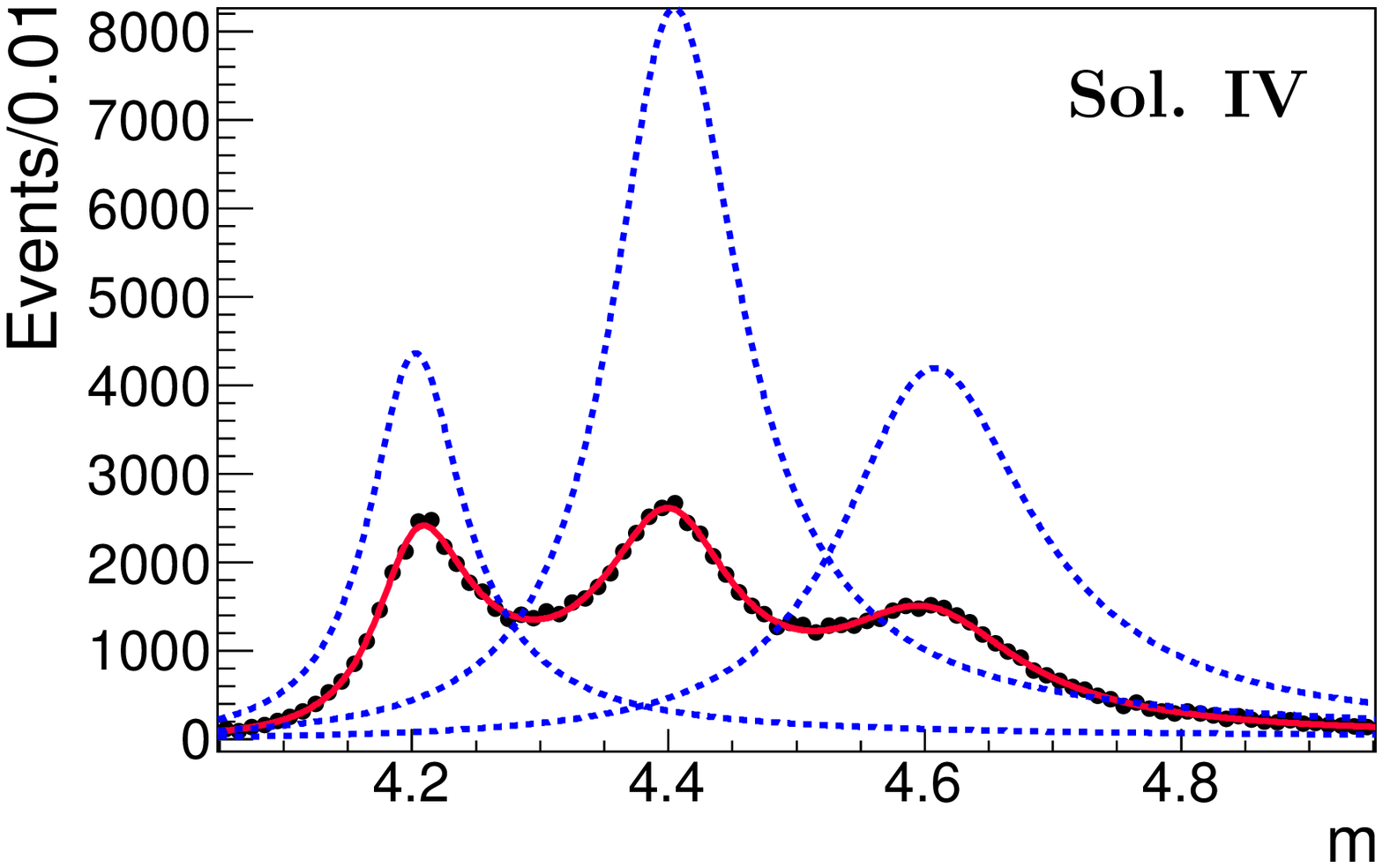}
\caption{The four solutions from the fit to the toy MC produced
mass spectra with the three interfering resonances included. The
solid curves show the best fit and the dashed curves show the
contributions from the three relativistic BW components.}
\label{fig2}
\end{figure}

According to the above probability density function and the first
set of input solution, a data sample of 100,000 events is generated
by using toy MC. Similarly, using the method mentioned earlier,
another three sets of solutions can be found numerically, which are exactly
repeated by fitting with the maximum likelihood method. The
comparison of the results is shown in Table~\ref{example22} and
the fitted plots are shown in Fig.~\ref{fig2}.

%}}}

\begin{table}[htbp]%{{{
\begin{center}
%\hspace{-15mm}
\caption[Example2]{Comparison between the extracted solution using
mathematical method and that from the fit
with three interfering relativistic BW functions.
A data sample of 100,000 events generated by toy MC
is used in the fit.}
\vspace{0.2cm} \label{example22}
\begin{tabular}{c|cc|cc|cc|cc}
\hline
 Item       & Sol. I (Input) & Fit I   & Sol. II & Fit II  & Sol. III & Fit III & Sol. IV  & Fit IV  \\
\hline
$\phi_f$   &  $\pi/2$   &1.57   & 2.63   & 2.63    &  3.44   &3.44     & 4.50    &4.50     \\
$\phi_h$   & $3\pi/4$   &2.36   & 6.14   & 6.14    &  5.12   &5.12     & 2.62    &2.62     \\
\hline
$d$        &1.00          &---    & 0.77   &---      & 0.45    &---      & 0.35    &---      \\
$\Ra$      &0.00          & 0.00  &-1.43   &-1.43    &-0.98    &-0.98    &-0.35    &-0.35    \\
$\Ia$      &1.00          & 1.00  & 0.80   & 0.80    &-0.30    &-0.30    &-1.63    &-1.63    \\
$\Rb$  &-$1/\sqrt{2}$  &-0.71  & 1.76   & 1.76    & 0.33    & 0.33    &-1.31    &-1.31    \\
$\Ib$   &$1/\sqrt{2}$  & 0.71  &-0.25   &-0.25    &-0.78    &-0.78    & 0.75    & 0.75    \\
\hline
$M_g$      &    4.20  &4.20   &  4.20  &4.20     &  4.20   &4.20     &  4.20   &4.20     \\
$\Gamma_g$ &    0.09  &0.09   &  0.09  &0.09     &  0.09   &0.09     &  0.09   &0.09     \\
$B_g\Gamma_{e^+e^-}^g$
           &    1.00  &1.03   &  1.29  &1.30     &  2.20   &2.21     &  2.85   &2.85     \\
$M_f$      &    4.40  &4.40   &  4.40  &4.40     &  4.40   &4.40     &  4.40   &4.40     \\
$\Gamma_f$ &    0.12  &0.12   &  0.12  &0.12     &  0.12   &0.12     &  0.12   &0.12     \\
$B_f\Gamma_{e^+e^-}^f$
           &    1.00  &1.02   &  3.46  &3.45     &  2.29   &2.28     &  7.94   &7.94     \\
$M_h$      &    4.60  &4.60   &  4.60  &4.60     &  4.60   &4.60     &  4.60   &4.60     \\
$\Gamma_h$ &    0.18  &0.18   &  0.18  &0.18     &  0.18   &0.18     &  0.18   &0.18 \\
$B_h\Gamma_{e^+e^-}^h$
           &    1.00  &1.01   &  4.07  &4.07     &  1.60   &1.60     &  6.53   &6.52     \\\hline
%\\\hline
%$B_g\cdot\Gamma_{\EE}$&  1&  &1.17022&        &1.4422   &         &2.49226  &         &3.06648  \\
%$B_f\cdot\Gamma_{\EE}$&  1&  &1.19941&        &3.88969  &         &2.6246   &         &8.58009  \\
%$B_h\cdot\Gamma_{\EE}$&  1&  &1.14942&        &4.52955  &         &1.78762  &         &6.99785
\end{tabular}
\end{center}
\end{table}%}}}

%}}}

\section{Discussion}%{{{
\label{sec:d}

As we found, when we need to describe a measured distribution using
three interfering resonances $|g(x)+z_{\alpha} f(x)+z_{\beta}h(x)|^2/d$
, $F(x)=f(x)/g(x)$ and $H(x)=h(x)/g(x)$ satisfy the relation of
Eq.~(\ref{eq:abcf}). If $f(x)$, $h(x)$, and $g(x)$ are widely used BW functions,
it has also been proved that such relation is exactly
satisfied.
In the case of three interfering resonances there occurred already
four equivalent solutions with the same likelihood function minimum.
Although the explicit analytical
formulae can not be derived between different solutions,
Eq.~(\ref{eq:bwary}) can be utilized to derive
the other three solutions numerically
from the solution obtained by fitting.
If three resonant amplitudes take simple or relativistic BW functions,
two data samples generated by toy MC are used
to cross check and verify our results.
For other complicated BW functions, the relations
Eqs.~(\ref{eq:abcf}), (\ref{eq:lirel}), (\ref{eq:dbw}), and
(\ref{eq:bwary}) still hold for $F(x)$ and $ H(x) $. And for other
forms of BW functions, with the coefficients obtained by
Eqs.~(\ref{eq:abcf}) and (\ref{eq:lirel}), the other solutions can
be derived numerically by using the method mentioned earlier.
The obtained numerical solutions agree well with those
from the fit, which justifies our method.
%We need to mention that when two data samples generated by toy MC as input
%are used to verify our method, only the central values are considered.
%In principle, one can obtain
%the errors of the parameters in the others solutions given the full
%covariant matrix is available for the first solution. But the relation
%between solutions is too complete to obtain the error transfer function.
%So the uncertainties of other solutions are not discussed, as the
%main purpose of this paper is to examine whether the other
%solutions exist or not, and how to find them.
We believe with the help of
finding other solutions numerically, it is easy to find all the solutions in real
fits to the experimental distribution as long as the initial values of resonant parameters
are set correctly.

%We also notice that for all solutions, the parameters of each
%resonance are same but the normalization factors. This implies that
%the couplings to decay channels are different for different solutions
%and some experimental reports may not be complete if only one solution
%was reported. So we suggest any experiment measurement with potential
%multiple-solution problem redo the analysis to find out the other solutions.
%Furthermore, only the sum of three coherent amplitudes
%is considered in this paper, the generalization to more amplitudes is
%still in progress.
%}}}

\acknowledgments

This work is supported in part by National Natural Science
Foundation of China (NSFC) under contract Nos. 11575017 and
11761141009; the Ministry of Science and Technology of China under
Contract No. 2015CB856701; and the CAS Center for Excellence in
Particle Physics (CCEPP).

%\begin{thebibliography}{7}%{{{
%\section*{Reference}

%\end{thebibliography}
\end{document}